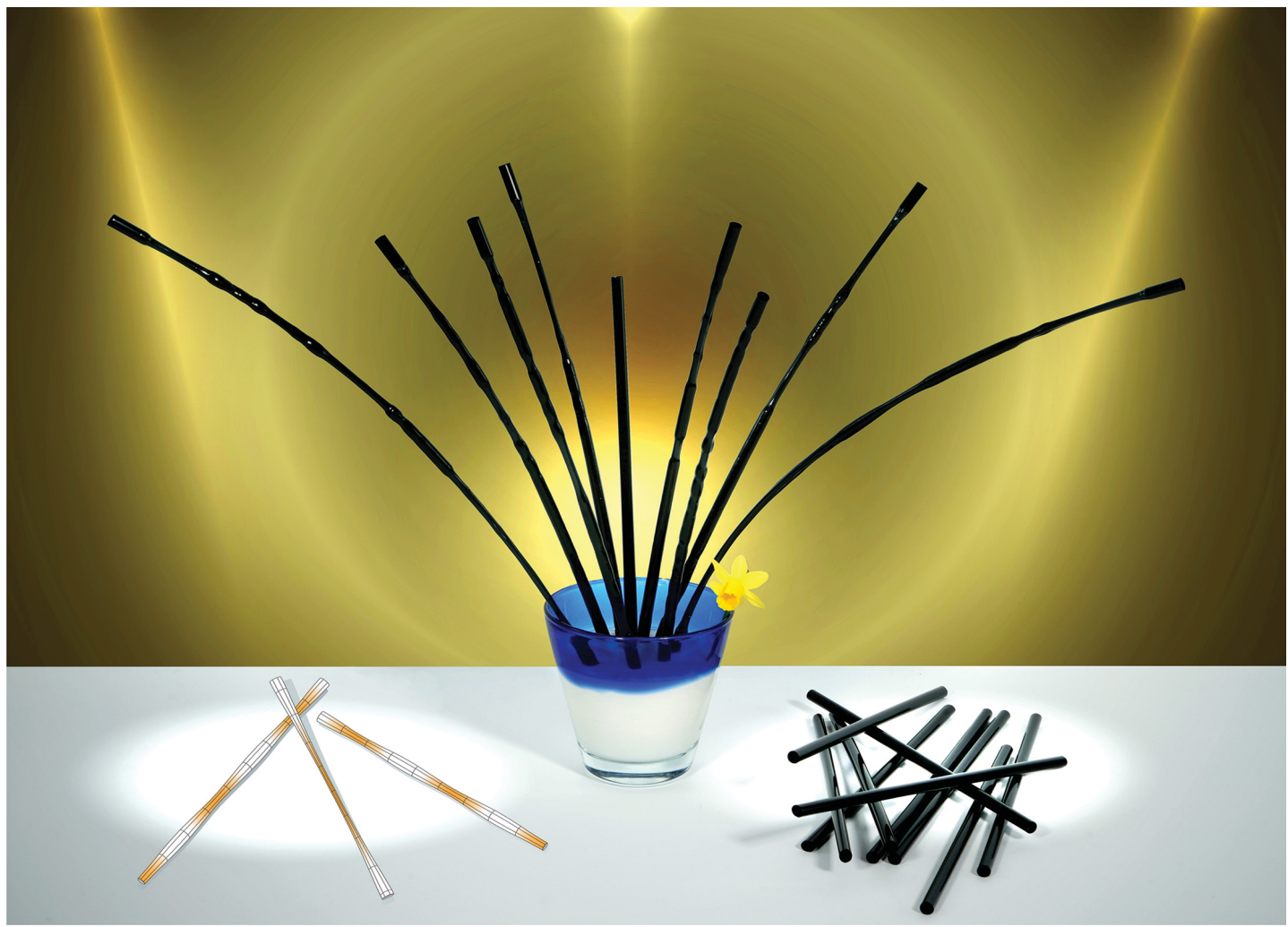

Showcasing research from Professor Davide Bigoni's "Instabilities lab", University of Trento, Via Mesiano 77 – 38123 Trento, Italy.

Necking of thin-walled cylinders *via* bifurcation of incompressible nonlinear elastic solids

Thin-walled tubes, made up of soft polypropylene, before and after tensile tests show multiple necking and formation of higher-order modes. These experimental findings were explained in terms of bifurcation for $J_2$-deformation theory of plasticity material. The theoretical framework leads to a series of new results, among which it is demonstrated that the classic Considère formula represents the limit of very thin tubes.

Photos of experiments were taken by Mr. M. Scandella and R. Springhetti, who also composed the image, under the supervision of D. Bigoni.

### As featured in:

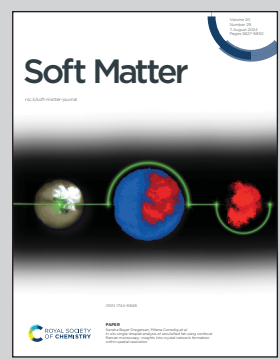

See Davide Bigoni *et al.*, *Soft Matter*, 2024, **20**, 5703.

ROYAL SOCIETY OF CHEMISTRY

rsc.li/soft-matter-journal

Registered charity number: 207890

# Necking of thin-walled cylinders *via* bifurcation of incompressible nonlinear elastic solids†

Roberta Springhetti, 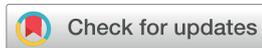 Gabriel Rossetto  and Davide Bigoni  *

Necking localization under quasi-static uniaxial tension is experimentally observed in ductile thin-walled cylindrical tubes, made of soft polypropylene. Necking nucleates at multiple locations along the tube and spreads throughout, involving the occurrence of higher-order modes, evidencing trefoil and fourth-foiled (but rarely even fifth-foiled) shaped cross-sections. No evidence of such a complicated necking occurrence and growth was found in other ductile materials for thin-walled cylinders under quasi-static loading. With the aim of modelling this phenomenon, as well as all other possible bifurcations, a two-dimensional formulation is introduced, in which only the mean surface of the tube is considered, paralleling the celebrated Flügge's treatment of axially-compressed cylindrical shells. This treatment is extended to include tension and a broad class of nonlinear-hyperelastic constitutive law for the material, which is also assumed to be *incompressible*. The theoretical framework leads to a number of new results, not only for tensile axial force (where necking is modelled and, as a particular case, the classic Considère formula is shown to represent the limit of very thin tubes), but also for compressive force, providing closed-form formulae for wrinkling (showing that a direct application of the Flügge equation can be incorrect) and for Euler buckling. It is shown that the $J_2$-deformation theory of plasticity (the simplest constitutive assumption to mimic through nonlinear elasticity the plastic branch of a material) captures multiple necking and occurrence of higher-order modes, so that experiments are explained. The presented results are important for several applications, ranging from aerospace and automotive engineering to the vascular mechanobiology, where a thin-walled tube (for instance an artery, or a catheter, or a stent) may become unstable not only in compression, but also in tension.

## 1 Introduction

Necking instability occurring in the course of a tensile test of a solid bar of circular cross-section was initially discovered in ductile metals and later detected in other materials, including nylon[1] and polymers.[2] Necking is a thoroughly analyzed problem; more than a century ago Armand Considère[3] related its initiation to the criterion of maximum load, which was later demonstrated to provide a lower bound (approached in the limit of infinitely slender specimens) to the critical load for bifurcation according to an axially-symmetric mode.[4–8] The development of neck curvature is accompanied by a strong alteration in the stress state, leading to triaxiality, as predicted by Bridgman.[9] Research on necking has developed along different lines, including the consideration of size effects through gradient models[10] and dynamic conditions.[11] Recently, models for the post-critical behaviour allowed a deep understanding of the phenomenon.[12,13]

*Instability Lab, University of Trento, via Mesiano 77, 38123-Trento, Italy.*
*E-mail: bigoni@ing.unitn.it; Tel: +39 0461 282507*

† Electronic supplementary information (ESI) available. See DOI: https://doi.org/10.1039/d4sm00463a

The present article originates from the experimental result shown in Fig. 1 and 2, reporting on a tensile test performed (in the 'Instabilities Lab' of the University of Trento, further details are deferred to Appendix A) on thin-walled tubes (generally used as drinking straws) made of polypropylene (detected through Fourier-transform infrared spectroscopy).

The material, characterized by an elastoplastic behaviour, is capable of sustaining axial stretches up to 6 without failure. The photo sequence shows that an initial necking localization occurs (in a random position along the tube) just after the peak is attained in the nominal stress/conventional strain curve, while multiple local necking phenomena progressively develop (without showing periodic or regular features) under increasing deformation of the tube, as shown in Fig. 3. Thinning of the cross-section, Fig. 5b, occurs while maintaining the plastic deformation incompressible, and it spreads along the specimen until the entire length is invaded, at which point the nominal stress starts to rise again. Interestingly, throughout this diffusive phase, when two consecutive necks localize at nearby positions, superior instability modes are observed to develop at increasing deformation in the intermediate region, as evident in Fig. 4 and 5c, with the cross-section assuming a trefoil shape, a fourth-foiled shape and, rarely, fifth-foiled shape. Multiple necking is common before dynamic fragmentation (a topic with a vast literature[14,15]), but rather unusual for quasi-static uniaxial tension. Moreover, necking for hollow rods was not reported so far (only a modest evidence of this phenomenon was found by us in moderately-thin metallic tubes, Fig. 12 in Appendix A) and, in particular, the fact that the tube is *thin-walled* represents a surprising circumstance from the mechanical point of view. In these conditions indeed, the stress triaxiality effect following necking in a solid bar is excluded, because plane stress continues to prevail even after neck formation.

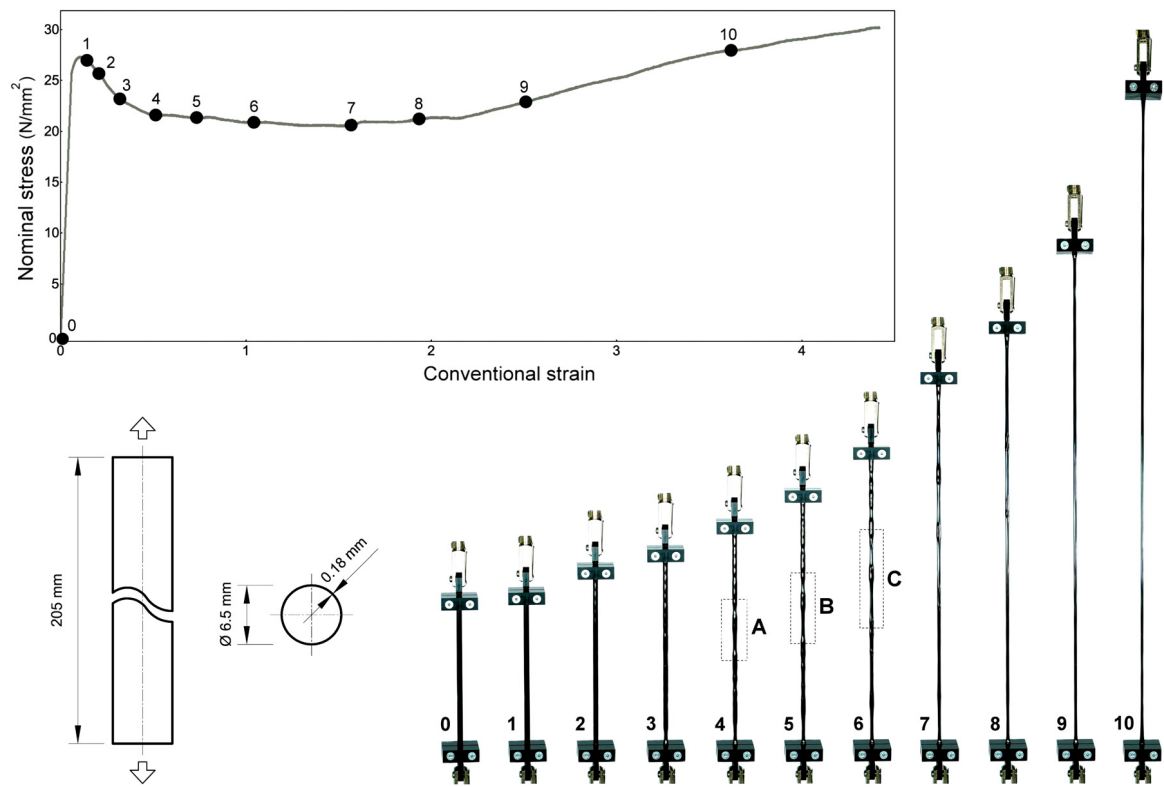

Fig. 1 Nominal stress vs. conventional strain obtained from a tensile test on a polypropylene (PP) thin-walled tube (6.5 mm initial diameter, 0.18 mm thickness, 205 mm length). During the test, the sample was brought up to a conventional strain of 4.5 at which the test was terminated before failure. The final thickness of the tube wall was 0.06 mm. Initiation and progression of necking is documented in Fig. 2.

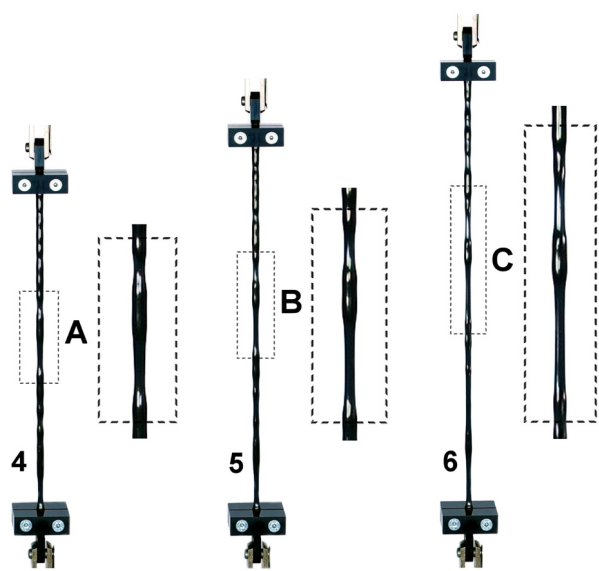

Fig. 2 Nucleation and development of multiple necks in polypropylene (PP) thin-walled tubes pulled in tension, detail of Fig. 1.

In response to the experimental evidence reported, an approach to bifurcation is introduced in the present article for a thin-walled cylinder, characterized by a constitutive equation belonging to a broad class of incompressible nonlinear elastic materials, including neo–Hookean and Mooney–Rivlin, and also the $J_2$-deformation theory of plasticity,[16,17] that imitates the loading behaviour of the analyzed tubes. In particular, for neo–Hookean, Mooney–Rivlin materials and also Ogden and Gent models, tensile bifurcations are ruled out, so that the results for these materials are included here only for reference when the axial force is compressive. On the other hand, the $J_2$-deformation theory of plasticity represents the simplest approach to tensile bifurcations and is shown to capture the observed necking patterns of the thin-walled tube. The latter constitutive equation mimics the in-loading branch of a plastic solid and allows the eventuality of tensile bifurcations. Remarkably, while the behaviour of cylindrical shells under *compression* and for compressible materials is a famous topic in mechanics,[18–20] the model of Flügge[21] does not allow tensile instabilities and was never extended neither to *tension* nor to incompressibility, which is the purpose of the present article. In particular, our theoretical approach retains the thinness of the tube wall through a condensation of the governing equations on the mean surface of the cylinder, thus defining a two-dimensional cylindrical shell. The developed theory allows analyzing the full spectrum of tensile and compressive bifurcations in cylindrical shells made of a non-linear elastic and incompressible material, never addressed so far. Our results, generalizing the Flügge outcomes in several ways, include the derivation of new closed-form solutions for wrinkling, for Euler buckling, and for necking. Under tension, it is analytically shown that the axial stretch for necking reduces to the famous Considère formula in the limit of very thin tubes. More in general, the modeling based on $J_2$-deformation theory of plasticity is shown not only to capture necking formation but also to explain the multiplicity of necks and the occurrence of higher-order modes. The presented results are important in a myriad of problems involving thin-walled tubes. These are well-known in civil, mechanical, and aerospace engineering, but are also of interest in the biomechanics of blood and urinary vessels, for the insertion and behaviour of stents and catheters. In all these examples, tubes are used in both compression and tension, and their instability represents a problem with serious consequences.

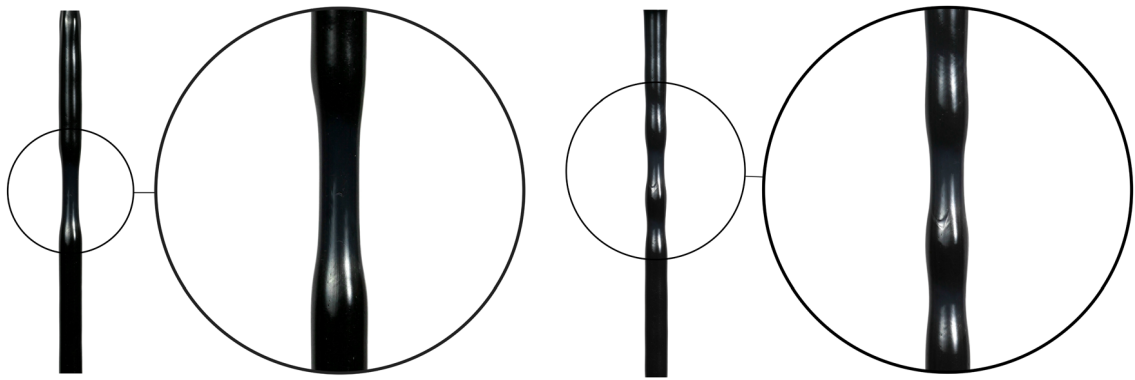

Fig. 3 Detail of necking development (on the left) and of the progressive formation of multiple necking in polypropylene (PP) thin-walled tubes subject to tension.

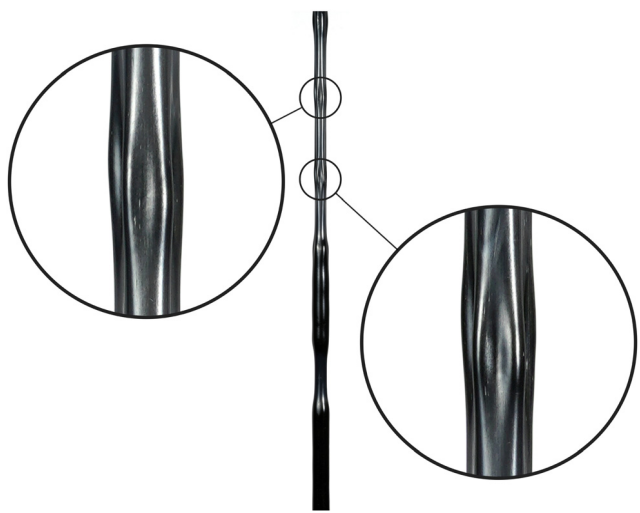

Fig. 4 Evidence of higher-order bifurcation modes in the necking of polypropylene (PP) thin-walled tubes under tension. Both trefoil and fourth-foiled modes have usually been observed, while, rarely, a fifth-foiled mode has also been found.

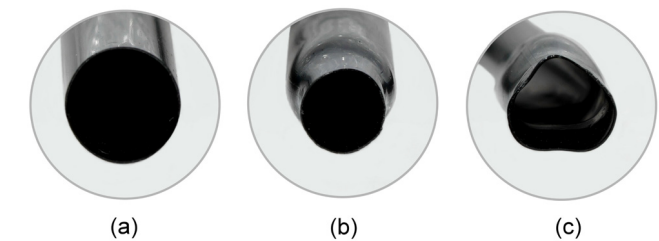

Fig. 5 Cross-sections of PP tubes under tension, showing: (a) the undeformed cross-section; (b) necking with uniform thinning; (c) necking involving a higher-order trefoil mode.

## 2 The pre-stress state and the incremental stress

A thin-walled, cylindrical shell of circular cross-section is considered, whose undeformed stress-free configuration is usefully described by means of cylindrical coordinates $(r_0, \theta_0, z_0)$, where the $z_0$-axis coincides with the axis of revolution of the shell. In undeformed configuration, the shell is characterized by length $L$, and by inner and outer radii $R_i$ and $R_e$, respectively, so that $a_0 = (R_i + R_e)/2$ denotes the initial mid-radius, while $t_0 = R_e - R_i$ represents the initial thickness. The fundamental path traversed by the structure prior to bifurcation is assumed to be a homogeneous, axisymmetric compression or tension in the longitudinal direction $z$, generated by a uniaxial Cauchy stress $\mathbf{T} = T_{zz}\mathbf{G}$, with $\mathbf{G} = \mathbf{e}_z \otimes \mathbf{e}_z$ ($\mathbf{e}_z$ denotes the unit vector aligned with the $z$-axis), therefore, in the current configuration

$$r = \lambda_r r_0, \quad \theta = \theta_0, \quad z = \lambda_z z_0.$$

Tensor $\mathbf{F} = \text{diag}\{\lambda_r, \lambda_\theta, \lambda_z\}$ denotes the deformation gradient, whose determinant $J = \det \mathbf{F} = 1$ for an incompressible material, corresponding to $\lambda_r \lambda_\theta \lambda_z = 1$. Maintaining the assumption of incompressibility and enforcing axial symmetry, $\lambda_r = \lambda_\theta$, the following relation is obtained

$$\lambda_r = \lambda_\theta = \lambda_z^{-1/2}.$$

A standard notation is used, with bold capital and lower case letters denoting tensors and vectors, respectively, and with capital or lower case letters indicating operators referred to material or spatial configurations, respectively. For incompressible materials, the Kirchhoff stress $\mathbf{K} = J\mathbf{T}$ (coincident with the Cauchy stress $\mathbf{T}$, as $J = 1$) can be written in the following form as a function of parameters $\alpha_1$ and $\alpha_{-1}$ depending on the constitutive law adopted:

$$\mathbf{K} = -\pi\mathbf{I} + \alpha_1\mathbf{B} + \alpha_{-1}\mathbf{B}^{-1}. \quad (2.1)$$

Here $\mathbf{B} = \mathbf{FF}^T = \text{diag}\{\lambda_r^2, \lambda_\theta^2, \lambda_z^2\} = \text{diag}\{\lambda_z^{-1}, \lambda_z^{-1}, \lambda_z^2\}$ is the left Cauchy-Green deformation tensor and $\pi$ is an arbitrary Lagrange multiplier, to be eliminated through the plane stress assumption in the pre-bifurcation stress, $K_{rr} = 0$ (note that for the initial pre-stress adopted, $K_{rr} = K_{\theta\theta} = 0$). Two incompressible materials are considered hereafter:[22]

• Mooney–Rivlin material, characterized by the strain energy function

$$W^{MR} = \frac{\mu_1}{2}(\lambda_r^2 + \lambda_\theta^2 + \lambda_z^2 - 3) - \frac{\mu_2}{2}(\lambda_r^{-2} + \lambda_\theta^{-2} + \lambda_z^{-2} - 3), \quad (2.2)$$

with the constants $\mu_1 > 0$ and $\mu_2 \leq 0$ ($\mu_2 = 0$ for the neo-Hookean material), such that $\mu = \mu_1 - \mu_2$ represents the shear modulus in the unloaded state. The relevant Cauchy stress parameters read as

$$\alpha_1 = \mu_1, \quad \alpha_{-1} = \mu_2; \quad (2.3)$$

• $J_2$-deformation theory of plasticity material,[16,17] defined by the strain energy function

$$W^{J_2} = \frac{K}{N+1}\epsilon_e^{N+1}, \quad (2.4)$$

with the constitutive stiffness parameter $K > 0$ and the hardening exponent $0 < N \leq 1$, while the effective strain

$$\epsilon_e = \sqrt{\frac{2}{3}\left[(\ln\lambda_r)^2 + (\ln\lambda_\theta)^2 + (\ln\lambda_z)^2\right]} \quad (2.5)$$

turns out to be $\epsilon_e = |\log\lambda_z|$ in the present context. The corresponding Cauchy stress parameters

$$\alpha_1 = \frac{K\epsilon_e^{N-1}\lambda_z[1 + \lambda_z^3(3\log\lambda_z - 1)]}{3(\lambda_z^3 - 1)^2},$$
$$\alpha_{-1} = \frac{K\epsilon_e^{N-1}\lambda_z^2[1 - \lambda_z^3 + 3\log\lambda_z]}{3(\lambda_z^3 - 1)^2} \quad (2.6)$$

depend on both the constitutive constants $K$ and $N$, as well as on the axial stretch $\lambda_z$.

The plane stress condition, requiring that the radial normal stress vanishes, $K_{rr} = 0$, allows the expression of the pre-stress components respectively to be

$$T_{zz}^{MR} = \frac{\lambda_z^3 - 1}{\lambda_z^2}(\mu_1\lambda_z - \mu_2), \quad T_{zz}^{J_2} = K|\log\lambda_z|^{N-1}\log\lambda_z. \quad (2.7)$$

The Oldroyd increment[23] of the Kirchhoff stress (and thus of the Cauchy stress in the case of incompressible materials), defined as

$$\overset{\circ}{\mathbf{K}} = \dot{p}\mathbf{I} + \mathbb{H}[\mathbf{D}]$$

with the Lagrange multiplier $\dot{p}$, can be written as

$$\overset{\circ}{\mathbf{K}} = \dot{p}\mathbf{I} + 2\pi\mathbf{D} + \left(\frac{d\alpha_1}{d\lambda_z}\mathbf{B} + \frac{d\alpha_{-1}}{d\lambda_z}\mathbf{B}^{-1}\right)D_{zz}\lambda_z \\ - 2\alpha_{-1}(\mathbf{B}^{-1}\mathbf{D} + \mathbf{D}\mathbf{B}^{-1}). \quad (2.8)$$

Multipliers $\pi$ and $\dot{p}$ are eliminated by imposing $K_{rr} = \overset{\circ}{K}_{rr} = 0$.

The incremental stress can be further simplified on the basis of the incremental incompressibility condition, prescribing $\text{tr}\,\mathbf{L} = 0$, namely,

$$D_{rr} = -(D_{\theta\theta} + D_{zz}). \quad (2.9)$$

The component $D_{rr}$ can therefore be replaced using condition (2.9) in the expression of the incremental stress $\overset{\circ}{\mathbf{K}}$; the final set of incremental equations for the two constitutive laws considered here are obtained by inserting eqn (2.3) and (2.6), respectively, into the following expressions:

$$\overset{\circ}{K}_{\theta\theta} = 2\lambda_z^{-1}(\alpha_1 - \alpha_{-1}\lambda_z^2)(2D_{\theta\theta} + D_{zz}),$$

$$\overset{\circ}{K}_{zz} = \lambda_z^{-2}\left[\lambda_z(\lambda_z^3 - 1)\left(\frac{d\alpha_1}{d\lambda_z}\lambda_z - \frac{d\alpha_{-1}}{d\lambda_z}\right)D_{zz} \right. \\ \left. + 2\alpha_1\lambda_z(D_{\theta\theta} + 2D_{zz}) - 2\alpha_{-1}(D_{\theta\theta}\lambda_z^3 + 2D_{zz})\right], \quad (2.10)$$

$$\overset{\circ}{K}_{r\theta} = 2\lambda_z^{-1}(\alpha_1 - \alpha_{-1}\lambda_z^2)D_{r\theta},$$

$$\overset{\circ}{K}_{rz} = 2\lambda_z^{-2}(\alpha_1\lambda_z - \alpha_{-1})D_{rz},$$

$$\overset{\circ}{K}_{\theta z} = 2\lambda_z^{-2}(\alpha_1\lambda_z - \alpha_{-1})D_{\theta z}.$$

# 3 Incremental deformations of an axially pre-stressed shell

The current configuration of the cylindrical shell is characterized by its length $l$, external and internal radii $r_e$ and $r_i$, respectively, and thus thickness $t = r_e - r_i$, mid-radius $a = (r_e + r_i)/2$, the latter defining the mid-surface, as well as the so-called 'reduced radius', $\bar{r} = r - a$, so that $-t/2 \leq \bar{r} \leq t/2$. The incremental kinematics of the thin-walled cylindrical shell is assumed in the form[24]

$$\mathbf{v}(\bar{r}, \theta, z) = \bar{\mathbf{v}}(\theta, z) + \bar{r}[\bar{\mathbf{n}}(\theta, z) - \mathbf{e}_r], \quad (3.1)$$

where $\bar{\mathbf{v}}(\theta, z)$ denotes the incremental displacement of the points along the current mid-surface of the shell, whose unit normal is

$$\bar{\mathbf{n}} \approx \mathbf{e}_r + (\bar{v}_\theta - \bar{v}_{r,\theta})/a\, \mathbf{e}_\theta - \bar{v}_{r,z}\, \mathbf{e}_z. \qquad (3.2)$$

The incremental displacements can therefore be detailed componentwise as

$$\begin{cases} v_r(\bar{r}, \theta, z) = \bar{v}_r(\theta, z), \\ v_\theta(\bar{r}, \theta, z) = \bar{v}_\theta(\theta, z) + (\bar{v}_\theta - \bar{v}_{r,\theta})\bar{r}/a, \\ v_z(\bar{r}, \theta, z) = \bar{v}_z(\theta, z) - \bar{v}_{r,z}\bar{r}. \end{cases} \qquad (3.3)$$

On the basis of the linearized kinematics in eqn (3.3), the components of gradient of incremental displacement $\mathbf{L} = \mathrm{grad}\,\mathbf{v}$ are

$$\begin{aligned}
L_{rr} &= 0, \\
L_{\theta\theta} &= [\bar{v}_r - (\bar{r}/a)\bar{v}_{r,\theta\theta} + (1 + \bar{r}/a)\bar{v}_{\theta,\theta}]/(a + \bar{r}), \\
L_{zz} &= -\bar{r}\bar{v}_{r,zz} + \bar{v}_{z,z}, \\
L_{\theta r} &= (-\bar{v}_{r,\theta} + \bar{v}_\theta)/a, \quad L_{zr} = -\bar{v}_{r,z}, \\
L_{r\theta} &= (\bar{v}_{r,\theta} - \bar{v}_\theta)/a, \quad L_{z\theta} = (-\bar{r}\bar{v}_{r,\theta z} + \bar{v}_{z,\theta})/(a + \bar{r}) \\
L_{rz} &= \bar{v}_{r,z}, \quad L_{\theta z} = -(\bar{r}/a)\bar{v}_{r,\theta z} + (1 + \bar{r}/a)\bar{v}_{\theta,z},
\end{aligned} \qquad (3.4)$$

so that the non-trivial components of the Eulerian incremental strain tensor $\mathbf{D} = (\mathbf{L} + \mathbf{L}^T)/2$ turn out to be

$$\begin{aligned}
D_{\theta\theta} &= [\bar{v}_r - (\bar{r}/a)\bar{v}_{r,\theta\theta} + (1 + \bar{r}/a)\bar{v}_{\theta,\theta}]/(a + \bar{r}), \\
D_{zz} &= -\bar{r}\bar{v}_{r,zz} + \bar{v}_{z,z}, \\
D_{\theta z} &= [-(\bar{r}/a)(2 + \bar{r}/a)\bar{v}_{r,\theta z} + (1 + \bar{r}/a)^2 \bar{v}_{\theta,z} \\
&\quad + \bar{v}_{z,\theta}/a]/[2(1 + \bar{r}/a)].
\end{aligned} \qquad (3.5)$$

When substituted into eqn (2.10), eqn (3.5) allow expressing the incremental stress $\overset{\circ}{\mathbf{K}}$ for both the constitutive laws considered here as functions of the constitutive parameters through eqn (2.3) and (2.6), respectively, the axial pre-stretch $\lambda_z$, and the components of the velocity along the mid-surface of the shell.

## 4 Generalized incremental stresses and equilibrium conditions

The incremental equilibrium of the pre-stressed shell within a relative Lagrangean description, with the current configuration assumed as reference (so that $\mathbf{F} = \mathbf{I}$), reads

$$\mathrm{div}\, \dot{\mathbf{S}} = 0, \qquad (4.1)$$

where the body forces are neglected. The increment of the first Piola–Kirchhoff stress tensor $\mathbf{S}$, denoted by $\dot{\mathbf{S}}$, has additionally to satisfy the traction-free surface boundary conditions on the lateral surface of the shell,

$$\dot{S}_{ir} = 0 \quad \text{as } \bar{r} = \pm t/2 \quad (i = r, \theta, z). \qquad (4.2)$$

In a relative Lagrangean description, the Oldroyd increment of the Kirchhoff stress is related to $\dot{\mathbf{S}}$ as

$$\overset{\circ}{\mathbf{K}} = \dot{\mathbf{S}} - \mathbf{L}\mathbf{K}, \qquad (4.3)$$

while the traction-free incremental boundary conditions, eqn (4.2), can be re-expressed through $\overset{\circ}{\mathbf{K}}$ as

$$\overset{\circ}{K}_{ir} = 0 \quad \text{as } \bar{r} = \pm t/2 \quad (i = r, \theta, z). \qquad (4.4)$$

According to the usual shell theory, a set of generalized stresses is introduced, representing the resultant forces and moments per unit length along the mid-surface of the shell,[24] as in Table 1. The relevant incremental stresses, based on the Oldroyd increment, are defined as

$$\begin{aligned}
\overset{\circ}{n}_{\cdot\theta} &= \int_{-t/2}^{t/2} \overset{\circ}{K}_{\cdot\theta}\, \mathrm{d}\bar{r}, \quad \overset{\circ}{n}_{\cdot z} = \int_{-t/2}^{t/2} \overset{\circ}{K}_{\cdot z}(1 + \bar{r}/a)\, \mathrm{d}\bar{r} \\
\overset{\circ}{m}_{\cdot\theta} &= -\int_{-t/2}^{t/2} \overset{\circ}{K}_{\cdot\theta}\bar{r}\, \mathrm{d}\bar{r}, \quad \overset{\circ}{m}_{\cdot z} = -\int_{-t/2}^{t/2} \overset{\circ}{K}_{\cdot z}\bar{r}(1 + \bar{r}/a)\, \mathrm{d}\bar{r},
\end{aligned} \qquad (4.5)$$

where $r, \theta, z$ are to be substituted in place of the subscript $\cdot$, while $K_{\cdot\theta}$ and $K_{\cdot z}$ represent the $\cdot\theta$ and the $\cdot z$ components of the Kirchhoff stress. Performing the integration of the incremental equilibrium equations across the thickness of the shell and the subsequent integration by parts with enforcement of the boundary conditions,[24] the incremental equilibrium can be expressed as

$$\begin{cases}
\overset{\circ}{m}_{\theta\theta,\theta\theta} + a\left(\overset{\circ}{m}_{\theta z} + \overset{\circ}{m}_{z\theta}\right)_{,\theta z} + a^2 \overset{\circ}{m}_{zz,zz} + a\overset{\circ}{n}_{\theta\theta} \\
\quad - \dfrac{Pa^2}{t}\int_{-t/2}^{t/2} \left(v_{\theta,\theta zz}\dfrac{\bar{r}}{a} + v_{z,zzz}\bar{r} + v_{r,zz}\right)\left(1 + \dfrac{\bar{r}}{a}\right)\mathrm{d}\bar{r} = 0, \\[4pt]
a\overset{\circ}{n}_{\theta\theta,\theta} + a^2\overset{\circ}{n}_{\theta z,z} - \overset{\circ}{m}_{\theta\theta,\theta} - a\overset{\circ}{m}_{\theta z,z} \\
\quad + \dfrac{Pa^2}{t}\int_{-t/2}^{t/2} v_{\theta,zz}\left(1 + \dfrac{\bar{r}}{a}\right)^2 \mathrm{d}\bar{r} = 0, \\[4pt]
a\overset{\circ}{n}_{zz,z} + \overset{\circ}{n}_{z\theta,\theta} + \dfrac{Pa}{t}\int_{-t/2}^{t/2} v_{z,zz}\left(1 + \dfrac{\bar{r}}{a}\right)\mathrm{d}\bar{r} = 0,
\end{cases} \qquad (4.6)$$

where $P = T_{zz}t$ is the pre-stress load per unit length along the mid-circular surface.

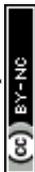

Table 1  The generalized variables adopted

| | | | |
|---|---|---|---|
| $n_{r\theta}$ | Radial shear force | $n_{rz}$ | Transverse shear force |
| $n_{\theta\theta}$ | Hoop force | $n_{\theta z}$ | Circumferential membrane shear force |
| $n_{z\theta}$ | Longitudinal membrane shear force | $n_{zz}$ | Longitudinal normal force |
| $m_{\theta\theta}$ | Hoop bending moment | $m_{\theta z}$ | Longitudinal twisting moment |
| $n_{z\theta}$ | Circumferential twisting moment | $m_{zz}$ | Circumferential bending moment |

## 5 The bifurcation problem

For each of the considered constitutive models, the final expressions for the Oldroyd increment of the Kirchhoff stress $\mathring{\mathbf{K}}$ based on eqn (2.10) are used for the computation of the generalized incremental stress variables involved in the average equilibrium eqn (4.6). The incremental displacements at bifurcation, based on the assumption of free sliding along perfectly smooth rigid constraints on the upper ($z = l$) and lower ($z = 0$) faces of the cylindrical shell (as commonly assumed[25]), are defined as

$$\begin{cases} \bar{v}_r(\theta, z) = c_1 \cos(n\theta) \cos(\eta z/a), \\ \bar{v}_\theta(\theta, z) = c_2 \sin(n\theta) \cos(\eta z/a), \\ \bar{v}_z(\theta, z) = c_3 \cos(n\theta) \sin(\eta z/a), \end{cases} \quad (5.1)$$

with the different modes selected by circumferential and longitudinal wave-numbers $n = 0, 1, 2,\ldots$ and $\eta = m\pi a/l$ ($m = 1, 2,\ldots$), respectively.

If vector $\mathbf{c}^T = \{c_1, c_2, c_3\}$ collects the amplitude of the bifurcation modes, a substitution of eqn (5.1) in the linearized kinematics, eqn (3.5), and in the average incremental equilibrium eqn (4.6), leads to the standard bifurcation condition

$$\mathbf{M}\,\mathbf{c} = 0, \quad (5.2)$$

requiring the singularity of matrix $\mathbf{M}$, namely,

$$\det \mathbf{M} = 0. \quad (5.3)$$

Eqn (5.3) only depends on the constitutive parameters, namely the ratio $\beta = \mu_2/\mu_1$ for the Mooney–Rivlin material and the hardening exponent $N$ for the $J_2$-deformation theory of plasticity, the circumferential and longitudinal wave-numbers $n$ and $\eta$, the axial stretch $\lambda_z$ and the dimensionless thickness of the shell $\tau = t/a = t_0/a_0$. The critical axial stretch $\lambda_z$ is computed from eqn (5.3) as a function of the wave-numbers $n$ and $\eta$, the geometrical variable $\tau$, and the constitutive parameters peculiar to the constitutive law adopted. The dimensionless load for bifurcation $p_z = -P/D$ (positive when compressive) can be finally obtained, where the parameter $D$ used to normalize the pre-stress is the product of the current thickness $t$ and a stiffness parameter, represented by the shear modulus in the unloaded state $\mu$ for the Mooney–Rivlin (or neo–Hookean) model ($D = \mu t$) or the stiffness parameter $K$ for the $J_2$-deformation theory of plasticity ($D = Kt$), respectively.

## 6 Bifurcation of axially-loaded incompressible thin-walled cylinders

Fig. 6 and 7 report the axial stretch at bifurcation for an axially-loaded thin-walled cylinder consisting of a Mooney–Rivlin and a $J_2$-deformation theory material, respectively. For completeness, both the cases of tensile and compressive axial stretch are considered, including, in addition the case of Mooney–Rivlin for comparison, even if in that case tensile bifurcations are excluded so that only compression is considered. The latter case will be shown to differ from the analogous treatment by Flügge and for this reason it is worth to be included. The axial stretch $\lambda_z$ is plotted as a function of the longitudinal wave-number $\eta$ for different geometrical ratios. Two very small thickness shells ($r_e/r_i$ = 1.01, $r_e/r_i$ = 1.02) have been analyzed, as well as a shell ($r_e/r_i$ = 1.06) corresponding to the tested PP tubes. Obviously, the thinner is the shell, the tighter to reality is the two-dimensional model. The figures also report the cross-section of the bifurcation modes. It is important to observe that for the $J_2$-deformation theory material pictured in Fig. 7, bifurcations are detected for both compression ($\lambda_z < 1$) and tension ($\lambda_z > 1$), while only bifurcation under compression ($\lambda_z < 1$) is foreseen for the Mooney–Rivlin material, Fig. 6. Moreover, while the loss of ellipticity with the potential outbreak of shear bands is *a priori* excluded for a Mooney–Rivlin material, it turns out to be possible for a $J_2$-deformation theory material when the boundary between the elliptic complex and the hyperbolic regimes is attained (*i.e.* for $N$ = 0.1, at $\lambda_z \simeq 0.448$ and $\lambda_z \simeq 2.234$ in compression and in tension, respectively). Therefore, Fig. 7 indicates that loss of ellipticity has no effects on the bifurcation envelope, because this condition is not attained for thin shells.

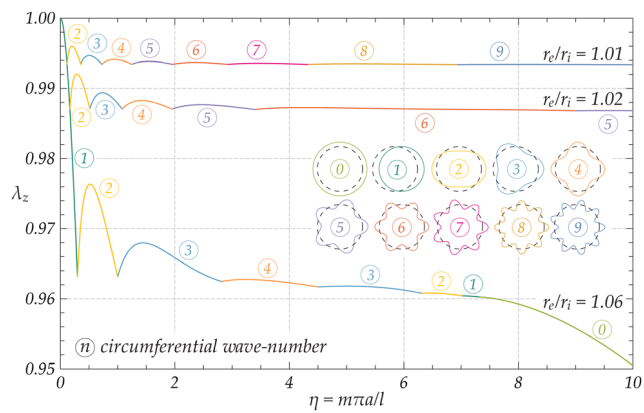

Fig. 6 Bifurcation upper envelopes for the critical stretch $\lambda_z$ of an axially-compressed thin-walled cylinder (geometrical ratios $r_e/r_i$ = 1.01, 1.02, 1.06) made up of a Mooney–Rivlin incompressible material with $\beta = \mu_2/\mu_1 = -0.1$. The curves corresponding to different values of the circumferential wave-number $n$ and the relevant bifurcation modes (plotted on the cross-section of the shell) are indicated with the symbol $\textcircled{n}$. All the bifurcations here occur in compression, tensile bifurcations were not detected indeed.

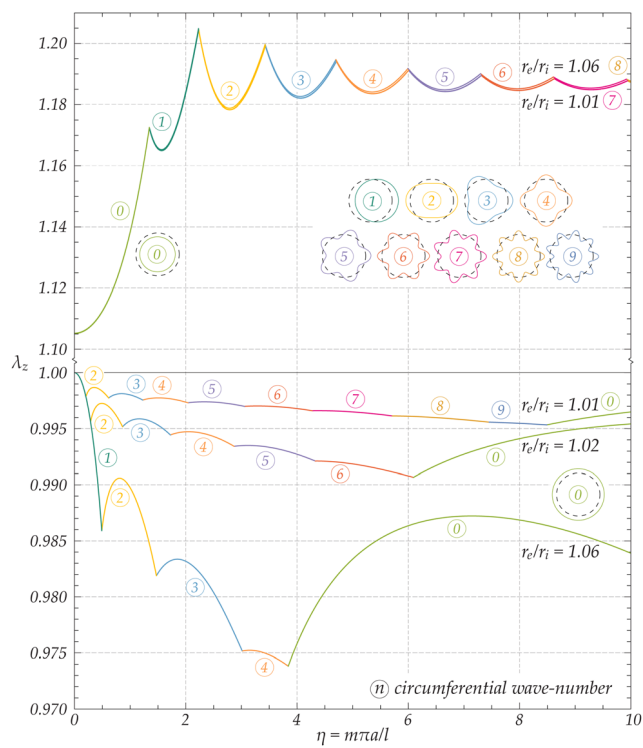

Fig. 7 Bifurcation envelopes for the critical stretch $\lambda_z$ of an axially-loaded thin-walled cylinder (geometrical ratios $r_e/r_i$ = 1.01, 1.02, and 1.06) made up of a $J_2$-deformation theory incompressible material with exponent $N$ = 0.1. The curves corresponding to different values of the circumferential wave-number $n$ and the relevant bifurcation modes (plotted on the cross-section of the shell) are indicated with the symbol $\textcircled{n}$. Bifurcations here occur both *in compression and tension*. The only modes differing in tension and compression are necking and bulging, corresponding to $n = 0$.

In compression ($\lambda_z < 1$), the mode $n = m = 1$ corresponds to Euler rod buckling and takes place when the shell is still almost undeformed, $\lambda_z \simeq 1$, for slender cylinders (with a slenderness ratio $L/a_0 \to \infty$). This is true for both the analyzed materials. Interestingly, only a number of the circumferential modes illustrated in Fig. 6 and 7 are involved in the bifurcation envelopes, according to the geometry of the thin-walled shell. For moderately large values of the longitudinal wave-number $\eta$ (according to the geometry of the shell), the cylinder undergoes longitudinal wrinkling with $n$ = 0, buckling into short longitudinal waves. Note that for the two very thin cylinders made up of the Mooney–Rivlin material pictured in Fig. 6, this mode is not evident within the range $0 < \eta < 10$, as it originates at larger values of $\eta$, here omitted for the sake of clarity.

For the $J_2$-deformation theory material pictured in Fig. 7, a variety of bifurcation modes becomes possible under both tension and compression, before failure of ellipticity. For $\lambda_z > 1$ the three reported shells behave very similarly, regardless of their geometric ratios. Over a threshold, $\lambda_z > \bar{\lambda}_z$, (whose value is asymptotically estimated in Section 7), in the limit $m \to 0$, a necking-type mode with $n$ = 0 occurs. The latter is a bifurcation mode compatible with the boundary conditions, characterized by a homogeneous transverse contraction and a sinusoidal variation along the longitudinal direction with a wavelength having the size of the cylinder length. Overall, the $J_2$-deformation theory material is found to correctly model necking in the previously reported PP tube, Fig. 1. After the occurrence of the first neck, the load tends to increase and this leads to neck formation in other parts of the sample. Finally, the appearance of higher-order modes is supported by the circumstance that the critical stretch for those does not exceed significantly ($\approx$10% for $N$ = 0.1) the critical stretch for the activation of necking. Therefore, after necks have invaded the entire sample a rise in load occurs, thus leading to the formation of higher-order modes.

In order to validate our generalization of the Flügge treatment for thin shells, a comparison between the proposed two-dimensional approach and the exact analysis (taken from Bigoni and Gei[22]) is illustrated in Fig. 8, reporting the lower bifurcation envelopes predicted for a $J_2$-deformation theory material in the case of a shell with a moderately thin wall, $r_e/r_i$ = 1.10, under tension. A modest difference is evident, while for thinner shells, the accuracy increases remarkably, so that the differences between the two approaches cannot be appreciated (as for instance when $r_e/r_i$ = 1.01). The results for the moderately thin shell reported in Fig. 8 show that the 2D-approach still allows a very precise prediction of necking, while the accuracy decreases when both the wave-numbers $n$ and $\eta$ grow.

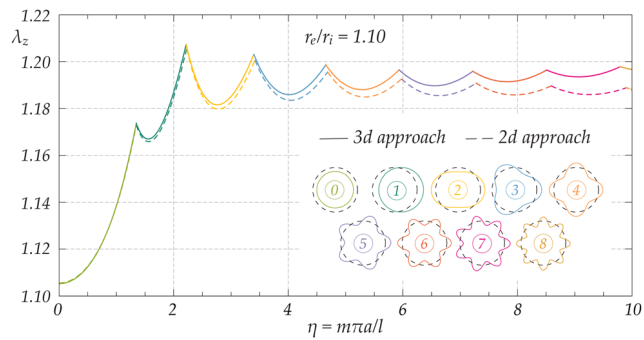

Fig. 8 Bifurcation lower envelopes for the critical stretch $\lambda_z$ of a moderately thin-walled cylinder (geometrical ratio $r_e/r_i$ = 1.10), made of a $J_2$-deformation theory incompressible material with exponent $N$ = 0.1 and subject to tension. The solid and dashed lines represent the predictions according to the 3d-exact analysis and the current Flügge 2d-approach, respectively, for different values of the circumferential wave-number $n$ corresponding to the different bifurcation modes indicated.

# 7 Asymptotic solutions for extreme cases

Extreme cases are considered, involving both tension and compression. For tension, necking is shown to occur at the Considère axial stretch when the thickness of the tube's wall tends to zero. Moreover, a closed-form analytical expression for the critical axial stretch for long-wavelength necking is provided. For compression, extreme cases are represented by wrinkling instability and Euler buckling of a rod.[24] For incompressible materials wrinkling leads to a new expression for its critical value: in particular, a new formula is provided for the Mooney–Rivlin material, for which Flügge's famous formula in the limit $\nu \to 1/2$ is not correct. Moreover, an expression for Euler buckling of a cylinder made up of $J_2$-deformation theory of plasticity is provided.

## 7.1 Necking instability for the $J_2$-deformation theory of plasticity

As pictured in Fig. 7, tensile instability is predicted on the basis of the $J_2$-deformation theory of plasticity, and its onset requires the attainment of a precise pre-stretch, representing the threshold for the activation of a long-wavelength necking, the mode with $n$ = 0 and $m \to 0$. In order to capture the asymptotic value for this axial pre-stretch, the determinant of matrix $\mathbf{M}$ in eqn (5.2) is evaluated for $n = 0$ and further expanded in a power series about $\eta = 0$ up to the 5th-order,

$$\det \mathbf{M}|_{n=0} = 2(36^3)\eta^4 \tau \log \lambda_z (\lambda_z^3 - 1)^5 (\lambda_z^3(\tau^2 + 12) + 2\tau^2)$$
$$\left[2(3\log \lambda_z - 3N - 1)\coth^{-1}\left(\frac{2}{\tau}\right) + \tau\right] + \mathcal{O}(\eta^6). \quad (7.1)$$

On enforcing vanishing of the above expression, and excluding unacceptable solutions, a closed-form formula for the necking onset of a thin-walled cylindrical shell made up of a $J_2$-deformation theory material is obtained

$$\bar{\lambda}_z = e^{N - \frac{\tau}{6\coth^{-1}\left(\frac{2}{\tau}\right)} + \frac{1}{3}}. \quad (7.2)$$

Eqn (7.2), never reported so far, shows that the critical stretch $\bar{\lambda}_z$ only depends on the constitutive exponent $N$ and on the dimensionless thickness of the shell $\tau$.

Actually, eqn (7.2) displays an interesting feature in the limit $\tau \to 0$, where the cylindrical shell becomes an 'ideal' shell, characterized by vanishing thickness. The limit yields exactly the celebrated Considère formula (for $J_2$-deformation theory)

$$\bar{\lambda}_z^0 = \lim_{\tau \to 0} \bar{\lambda}_z = e^N, \quad (7.3)$$

showing that *an ideal shell still admits a non-null value of critical stretch*, $\bar{\lambda}_z^0 = e^N$, *for necking instability*. In the case $N = 0.1$, the shells reported in Fig. 7, characterized by $r_e/r_i = \{1.06, 1.02, 1.01\}$ and $\tau = \{0.05825, 0.0198, 0.00995\}$, are well represented by the ideal shell approximation, with critical stretches $\bar{\lambda}_z = \{1.10528, 1.10518, 1.10517\}$ tending to the critical stretch of the ideal shell, $\bar{\lambda}_z^0 \approx 1.10517$.

### 7.2 Limit cases for bifurcations in compression

**Longitudinal wrinkling.** For mildly long cylindrical shells made up of a linear elastic isotropic incompressible ($\nu \to 0.5$) material, the well-known solution by Flügge[21] in terms of dimensionless pressure,

$$p_{z,\text{Flügge}}^w = \frac{P}{\mu t} = 2\tau, \quad (7.4)$$

can be rigorously obtained on the basis of the proposed two-dimensional approach for a material with neo–Hookean constitutive law. However, eqn (7.4) is shown to be *incorrect* for a Mooney–Rivlin material, so that the following new expression

$$p_z^{MRw} = 2(1-\beta)\tau + \frac{17\beta - 41}{6}\tau^2 + \mathcal{O}(\tau^3), \quad (7.5)$$

is obtained below. Eqn (7.5) provides the critical axial load for longitudinal wrinkling of a Mooney–Rivlin material and reduces to eqn (7.4) at the first-order when $\beta = 0$.

Eqn (7.5) can be obtained as follows. Within an intermediate range of the wave-numbers $\eta$, the bifurcation load turns out to be almost independent of both wave-numbers $n$ and $m$, as visible in the lower curve in Fig. 9, representing the dimensionless line force $p_z^{MR}$ for bifurcation of a Mooney–Rivlin material with $\beta = -0.1$. The curve corresponding to the neo–Hookean material was omitted in the figure, because only slightly below but almost superimposed to the curve corresponding to the Mooney–Rivlin material, in agreement with eqn (7.5). In order to capture this flat portion of the curve, the critical axial stretch $\lambda_z$ is approximated by enforcing the singularity of matrix $\mathbf{M}$ in eqn (5.2), first computed for null circumferential wave-number ($n = 0$), then linearly expanded into a Taylor series about $\lambda_z = 1$. The minimum of function $p_z(\eta)$ with respect to the longitudinal wave-number $\eta$ is sought by imposing the stationarity of function $\lambda_z(\eta)$, being $p_z(\lambda_z)$ a monotonic function. Five solutions are found from the latter condition: one is trivial, two are imaginary conjugates and two are reals with opposite signs. The only admissible solution is selected and replaced in the expression of $p_z(\eta)$, successively developed as a Taylor series about $\tau = 0$, to return eqn (7.5).

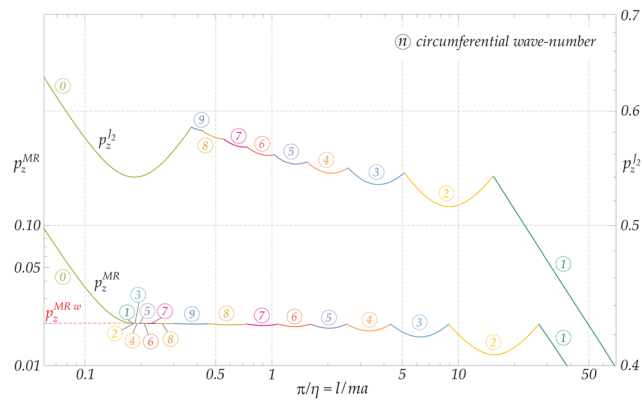

Fig. 9 Bi-logarithmic representation (two different vertical scales are used) of the lower envelopes of the dimensionless load, $p_z^{J_2}$ (for the $J_2$-deformation theory material with $N = 0.1$) and $p_z^{MR}$ (for the Mooney–Rivlin material with $\beta = -0.1$), vs. $\pi/\eta$, for the axially-compressed thin-walled cylinder with $r_e/r_i = 1.01$. For moderate values of $\pi/\eta$, the presence of a horizontal flat portion of the curve (highlighted as a red dashed line), corresponding to longitudinal wrinkling and leading to eqn (7.5), is evident for the Mooney–Rivlin material (and also for the neo–Hookean, not reported as almost superimposed to the Mooney–Rivlin material), but absent for the $J_2$-material.

Finally, Fig. 9 reports also the bifurcation curve for pressure $p_z^{J_2}$ concerning the $J_2$-deformation theory material. Differently from the Mooney–Rivlin material ($p_z^{MR}$), the flat portion of the curve leading to the longitudinal wrinkling *is not present for the $J_2$-material*, so that an equation analogous to eqn (7.5) cannot be obtained for this material.

**Euler rod buckling.** The anti-symmetric mode of bifurcation with $m = n = 1$ corresponds to the Euler buckling of a rod constrained by sliding clamps at both ends. The fundamental geometric parameter governing buckling is the stubbiness ratio, measured in the undeformed configuration, $\alpha_0 = a_0/L$. The latter is related to the analogous parameter defined in the current configuration as $\alpha = a/l = \lambda_r a_0/\lambda_z L = \lambda_z^{-3/2}\alpha_0$. The asymptotic expression of the buckling load for both Mooney–Rivlin and $J_2$-deformation theory materials is derived below on the basis of a perturbative approach,[24,26] whose general procedure is briefly outlined below. Upon generating a series expansion of order $I$ for the axial stretch around the unstretched configuration as $\lambda_z = 1 + \lambda_{zi}\alpha^i$, $1 \leq i \leq I$, the expression taken by the determinant of matrix **M** in eqn (5.2) with the approximation for $\lambda_z$ above is first computed for both $m = 1$ and $n = 1$, and then expanded into a Taylor series about $\alpha = 0$ (slender columns) up to order $I + 2$. Enforcing the vanishing of the coefficients for all the exponents of $\alpha$ in the expansion of $\det \mathbf{M}|_{m=n=1}$, the approximation for $\lambda_z$ is made explicit, and turns out to involve only even powers of $\alpha$. The condition $\alpha^2 \lambda_z^3 = \alpha_0^2$ is used at this stage to connect the current and initial stubbiness ratios, by introducing the obtained approximation for $\lambda_z$, together with the approximation $\alpha = \alpha_{0j}\alpha_0^j$, $0 \leq j \leq J$ (a posteriori, only odd powers are found to be different from zero). Enforcing that all the exponents of $\alpha_0$ have null coefficients, the asymptotic solution for the critical axial stretch $\lambda_z$ as a function of $\alpha_0$ is finally obtained and substituted into the expression for the current longitudinal force (positive when compressive) on the cylinder:

$$N_z = -\pi(r_e^2 - r_i^2)T_{zz} = -\pi\lambda_z^{-1}(R_e^2 - R_i^2)T_{zz}. \quad (7.6)$$

In this way the well-known critical axial force for Euler buckling is derived for the two considered constitutive models.

• For the Mooney–Rivlin model, the result provided by eqn (7.6), further expanded into a power series about $\alpha_0 = 0$ up to the 4th-order, is

$$N_z^{\text{MR}} = \frac{\pi^3}{144}\mu a_0^2 \tau (432 + 120\tau^2 - \tau^4)\alpha_0^2 + \mathcal{O}(\alpha_0^4) \quad (7.7)$$

Remarkably, up to the considered order of approximation, the asymptotic solution is independent of the constitutive parameter $\beta$. Therefore, the Euler buckling loads for the neo–Hookean and the Mooney–Rivlin *incompressible* materials turn out to have the same asymptotic expressions. Moreover, the solution in eqn (7.7) coincides with the corresponding equation [eqn (6.20) in the ref. 24] for a *compressible* neo–Hookean material in the limit as $\nu \to 1/2$.

• For the $J_2$-deformation theory of plasticity, the simplest form for the leading term of the solution, eqn (7.6), is obtained using second-order approximations for both the axial stretch $\lambda_z$ and $\alpha(\alpha_0)$. The expression never obtained so far,

$$N_z^{J_2} \approx 2\pi K a_0^2 \tau \frac{(-\log \zeta)^N}{\zeta} \quad (7.8)$$

with $\zeta = 1 + \frac{\pi^2}{864}\alpha_0^2 [\tau^4 - 12(1 + 9N)\tau^2 - 432N]$, provides a very good approximation for the Euler buckling solution for $\alpha_0 \to 0$.

## 8 Conclusions

Quasi-static tensile tests on thin-walled tubes made of polypropylene show new features, involving necking nucleation and growth, multiple neck formation, and occurrence of higher-order bifurcation modes, with trefoil or fourth-foiled (and, rarely, even fifth-foiled) transverse shape. The necking develops during a highly-ductile plastic flow and surprises because the tube wall is thin, so that stress triaxiality does not follow instability. Such a peculiar behaviour was not found by us, at least up to the observed extent, on any other material.

To model these experimental findings, the celebrated bifurcation theory for compressed cylinders proposed by Flügge has been generalized to include tension, incompressibility, and a broad class of nonlinear elastic constitutive laws, which may also describe the plastic branch of ductile materials, the so-called $J_2$-deformation theory of plasticity. The theoretical framework has revealed a number of new results valid for thin-walled cylindrical shells: (i) in tension, closed-form formulas for necking, revealing that the Considère formula corresponds to the 'ideal shell' limit; in compression, (ii) wrinkling of a Mooney–Rivlin material, thus improving and extending a famous result from Flügge, and (iii) Euler buckling for Mooney–Rivlin and $J_2$-deformation theory of plasticity materials.

The $J_2$-deformation theory of plasticity has been shown to correctly model the onset of the observed necking in PP tubes, the formation of multiple necking, and the occurrence of higher-order modes, with the latter two effects related to the 'flatness' of the bifurcation curve in tension.

## Author contributions

D. B.: conceptualization, formal analysis, funding acquisition, investigation, methodology, project administration, resources, supervision, validation, visualization, writing-original draft, writing-review and editing, and cover artwork-review; R. S.: conceptualization, formal analysis, investigation, methodology, resources, software, supervision, validation, visualization, writing-original draft, writing-review and editing, composition and realisation of the cover artwork; G. R.: conceptualization, formal analysis, investigation, visualization, and writing-original draft.

## Data availability

The authors do not have other data available in addition to the submitted material.

## Conflicts of interest

There are no conflicts of interest to declare.

## Appendix

### A Experimental evidence of necking in tubes

Several tensile tests were performed directly on PP and metal tubes, while three tests were executed on dog-bone samples (cut out after sectioning the PP tubes along a generatrix), using an electromechanical testing machine (Messphysik Midi 10 for the PP tubes and Messphysik Beta 100 for the metal tubes) under displacement control at a speed of 0.5 mm s$^{-1}$. The load was measured by means of a REP TC4 (RC 10 kN) and a REP TC4 (RC 100 kN) loading cells, and the displacement of the crossbar of the testing machine was recorded.

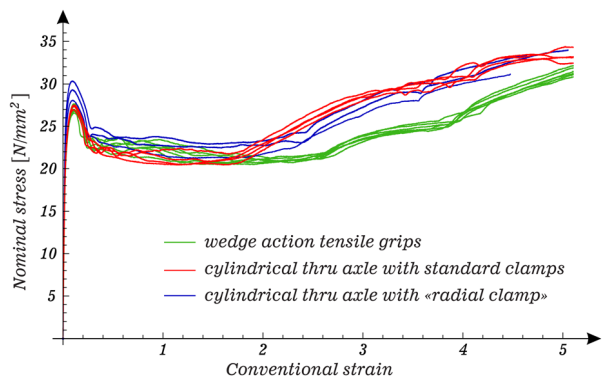

Fig. 10 Nominal stress vs. conventional strain data for tensile tests on polypropylene (PP) thin-walled tubes (6.5 mm initial diameter, 0.18 mm thickness, 205 mm length). With a couple of exceptions, the samples were brought up to a conventional strain of approximately 5 where the test was terminated before failure. Three different gripping devices have been tested and the relevant experimental data are reported using the same colour. The 'radial clamp' was designed by us and is shown in Fig. 11.

The results from a dozen of tests on PP tubes (all characterized by initial dimensions: 6.5 mm initial diameter, 0.18 mm thickness, and 205 mm length) are reported in Fig. 10, in terms of nominal stress vs. conventional strain curves. Three tests on dog-bone samples provided results in agreement with those performed on tubes. However, the flattening of the samples after excision from a cut tube induces a strong initial stress state, so that these tests were not continued and are not reported. Overall, the experimental results suggested the use of $N = 0.1$ in the reported numerical examples for bifurcation.

The geometry of the PP tubes was carefully investigated in terms of variation in diameter and thickness, to reveal fabrication defects and to select the 'best' samples for subsequent testing.

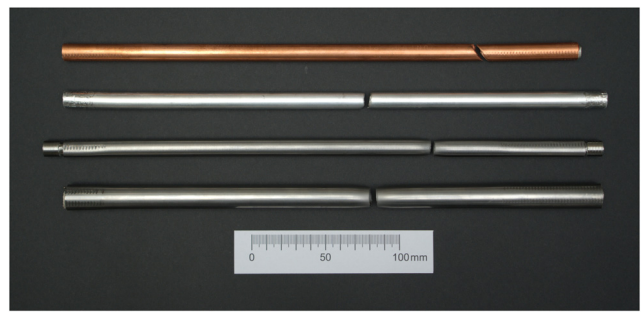

Fig. 12 Photographs taken after the failure of metallic tubes subject to a tensile test. From the upper to the lower side: copper and aluminium tubes ($R_i$ = 6 mm, and $R_e/R_i$ = 1.2) and two steel tubes ($R_i$ = 6 mm, $R_e/R_i$ = 1.25, and $R_i$ = 8 mm, $R_e/R_i$ = 1.23). While the steel tubes show an initial development of necking, the latter is not observed in neither copper nor aluminium tubes. Multiple necking and development of higher modes were not found. These metallic samples behaved completely differently from the PP tubes.

The variability in diameter (minimum and maximum values) was evaluated at 5 equally-spaced sections along the tube for 31 samples, using a Palmer caliper (accuracy ±0.002 mm). The extreme minimum and maximum diameters were found to be 6.02 mm and 7.35 mm, respectively. Along the length of each sample, standard deviation for all measured diameters (10 data) fluctuates between 0.02 mm and 0.57 mm (note that this parameter was found to be smaller than 0.1 mm in 11 samples). The thickness, detected at 13 different sections on 7 samples cut along a generatrix, exhibits a much smaller variability, with its average value being 0.18 mm.

While tests on the metal tubes were standard, the experimental setup for the tests on PP tubes involved a non-standard feature shown in Fig. 11, namely, the tensile grip of the PP tube. Three alternatives have been tested for the constraints at the tube ends: (i) standard wedge action tensile grips (from Instron), which squeezed the ends of the tube, therby triggering the initiation of necking; (ii) the wedge action tensile grips, in which a tight-fitting thru axle was inserted to prevent the squeezing of the tube, and (iii) a 'radial clamp' device (detailed in Fig. 11 on the right) fixed against the entire external surface of the tube ends, in which a tight-fitting thru axle was inserted. Fig. 10 clearly shows that results in terms of nominal stress/conventional strain are only weakly affected by the different clamping conditions. The same modest sensitivity to end conditions was observed for multiple necking and development of higher-order modes. These features are shown for two tests performed with the 'radial clamp' in Fig. 1 and also in Fig. 13, added for completeness. Videos of experiments are available in the ESI.† It is worth mentioning that the stress/strain curves reported in Fig. 1, 10 and 13 look very similar to the ones obtained for a polycarbonate compact rod under tension at different strain rates.[27]

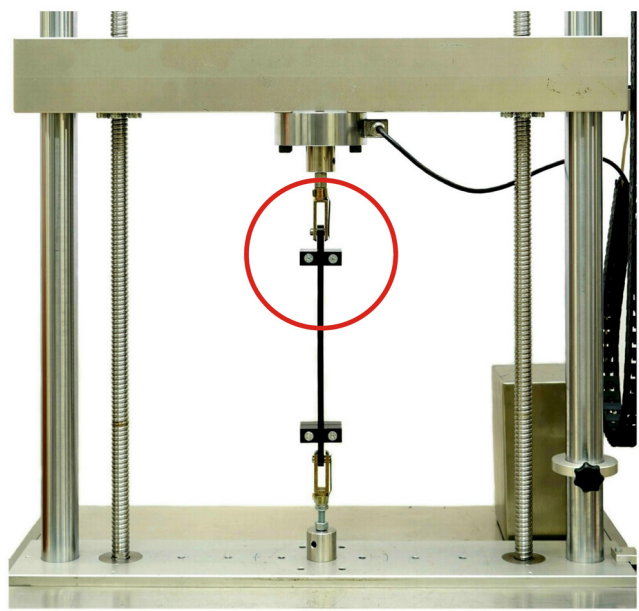
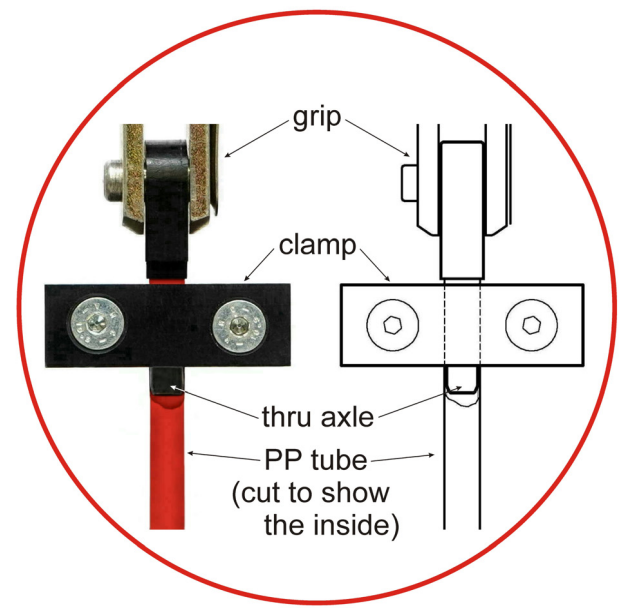

Fig. 11 The experimental set-up for tensile tests on polypropylene (PP) thin-walled tubes. It is based on an electromechanical testing machine (Messphysik Midi 10). The 'radial clamp' device is shown on the right.

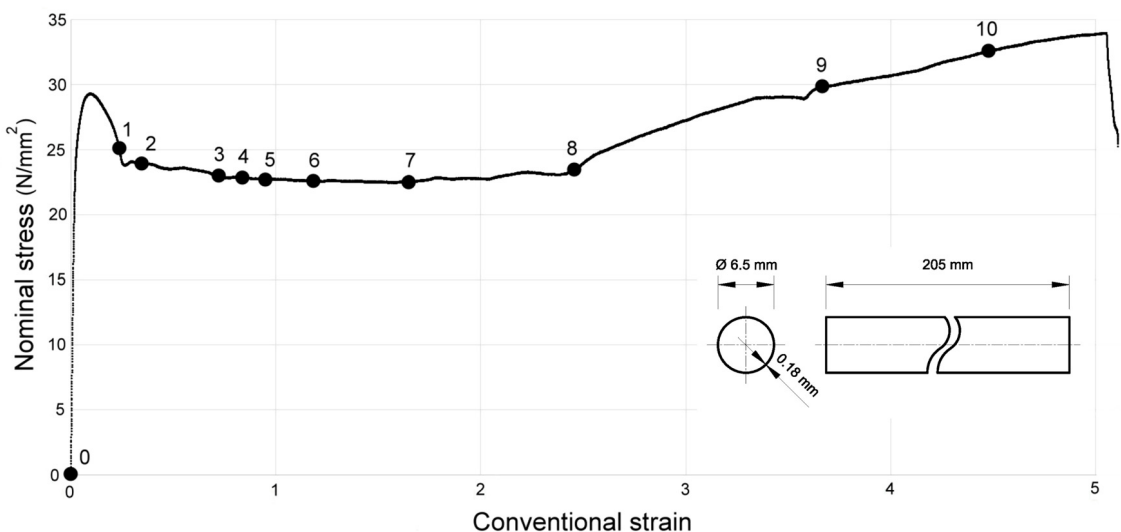

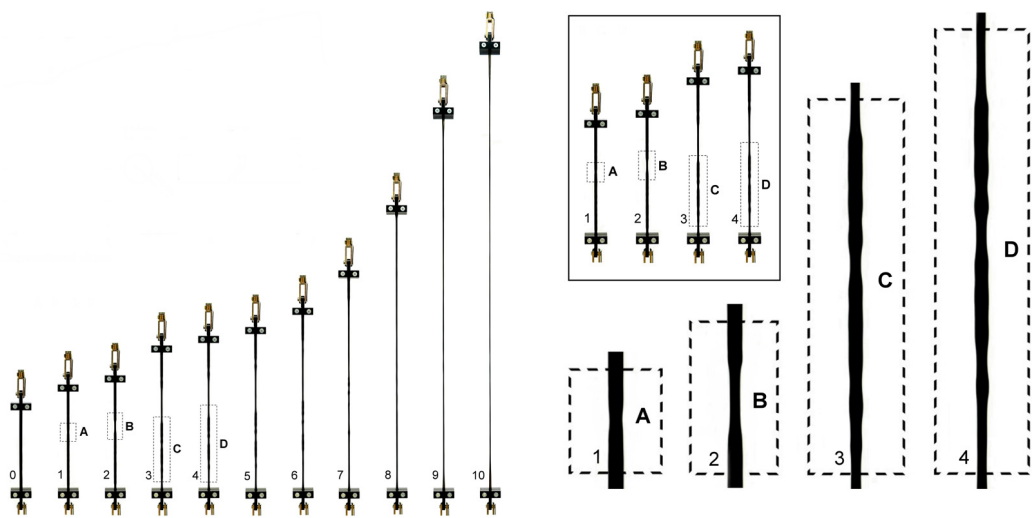

Fig. 13 Upper part: Nominal stress vs. conventional strain from a tensile test on a polypropylene (PP) thin-walled tube. Through 'radial clamps', the sample was brought before failure up to the conventional strain of 5. The final thickness was 0.06 mm. Lower part: nucleation and development of multiple necking. Although hardly visible in the photographs, higher-order modes were observed.

In order to confirm the peculiar behaviour of PP tubes, evidence of necking was investigated in metal tubes made up of copper and aluminium (initial dimensions $R_i$ = 6 mm, $R_e/R_i$ = 1.2) and steel (two samples having initial dimensions $R_i$ = 6 mm, $R_e/R_i$ = 1.25, and $R_i$ = 8 mm, $R_e/R_i$ = 1.23). Photographs taken after failure under uniaxial tension are reported in Fig. 12, which show that a moderate necking was only observed in the steel tubes (which are also thicker than the others), while multiple necking or higher-order modes were not found.

## Acknowledgements

D. B. and G. R. acknowledge funding from the European Research Council (ERC) under the European Union's Horizon 2020 research and innovation programme, Grant agreement No. ERC-ADG-2021-101052956-BEYOND. R. S. gratefully acknowledges funding from the European Union, Grant agreement No. ERC-CoG-2022-SFOAM-101086644. The authors thank Mr. M. Scandella and Mr. G. De Sero for their contribution to the realization of the cover artwork and of the experiments at the 'Instabilities Lab' of the University of Trento.

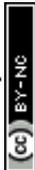

## Notes and references